\documentclass[aps,prb,reprint,twocolumn,superscriptaddress,showpacs,floatfix,longbibliography]{revtex4-1}

\usepackage{graphicx}
\usepackage{amsmath}
\usepackage{amssymb}
\usepackage{bbm}
\usepackage{color}
\usepackage{enumitem}
\usepackage{nicefrac}
\usepackage{multirow}
\usepackage{braket}
\usepackage{mathrsfs}

\usepackage[colorlinks,citecolor=blue,urlcolor=blue,bookmarks=false,hypertexnames=true]{hyperref} 

\begin{document}
\title{Can second-order perturbation theory accurately predict electron density of open-shell molecules? The importance of self-consistency}
\date{\today}
\author{Lan Nguyen Tran}
\email{tnlan@iami.vast.vn}
\affiliation{HCMC Institute of Physics, National Institute of Applied Mechanics and Informatics, Vietnam Academy of Science and Technology, Ho Chi Minh City 700000, Vietnam}

\begin{abstract}
Electron density distribution plays an essential role in predicting molecular properties. It is also a simple observable from which machine-learning models for molecular electronic structure can be derived. In the present work, we present the performance of the one-body M{\o}ller-Plesset second-order perturbation (OBMP2) theory that we have recently developed. In OBMP2, an effective one-body Hamiltonian including dynamic correlation at the MP2 level is derived using the canonical transformation followed by the cumulant approximation. We evaluate electron density and related properties of three groups of open-shell systems: atoms and their ions, main-group radicals, and halogen dimmers. We find that OBMP2 outperforms standard MP2 and density functional theory in all cases considered here, and its accuracy is comparable to coupled-cluster singles and doubles (CCSD), a higher-level method. OBMP2 is thus believed to be an effective method for predicting accurate electron density of open-shell molecules.
\end{abstract}

\maketitle

\section{Introduction}
Molecular properties such as electrostatic moments (e.g., charges, dipole, quadrupoles), electrostatic potentials, and electrostatic interaction energies can be derived from electron density distribution. Information needed to characterize chemical systems can be thus encoded in electron density \cite{parr1980density, fabrizio2019electron}. Particularly, electron density distribution and its derivatives are among the simplest and generally useful tools to investigate non-covalent compounds \cite{fabrizio2019electron, kumar2021towards,boto2020nciplot4}. Recently, several works have developed density-learning models whose parameters are trained by fitting to reference electron densities obtained from {\it ab initio} quantum chemical calculations \cite{fabrizio2019electron,cuevas2020analytical}. To achieve a good model,  one needs to prepare many high-quality training data. Low-scaling methods able to generate accurate electron density are thus highly desirable. 

M{\o}ller-Plesset Second-order perturbation theory (MP2) is the simplest correlated wavefunction method beyond Hartree-Fock (HF). In MP2, the correlated density matrix is evaluated using the double-excitation MP2 amplitude that relies on HF molecular orbitals and orbital energies. In many cases, particularly for open-shell systems where unrestricted Hartree-Fock (UHF) solutions severely suffer from spin-contamination, correlated MP2 density is not good \cite{OOMP2-HFCC-2010, tran2021improving}. One can relax MP2 density by solving $Z$-vector equation for the orbital response. However, as shown previously, the singularity of the inverse of the orbital Hessian matrix causes the violation of $N$-representability of relaxed density, leading to unphysical discontinuities \cite{MP2-Molphys2009}. 

Orbital-optimized MP2 (OOMP2) and its spin-scaled variants have been developed actively\cite{OOMP2-HeadGordon,OOMP2-Neese,OOMP2-Sherrill,Bozkaya2013-GradOMP2,Bozkaya2014-OMP2,Bozkaya2014-GradOMP2,OOMP2-JCP2013,OOMP2-Molphys2015,OOMP2-Molphys2017,OOMP2-JCTC2018}. In OOMP2, orbitals are optimized by minimizing the Hylleraas functional or its generalization. OOMP2 can eliminate the spin contamination present in UHF for open-shell molecules \cite{Bozkaya2014-GradOMP2,OOMP2-HFCC-2010}. OOMP2 has been shown to outperform standard MP2 for many energy-related properties. However, to our knowledge, we are only aware of a few studies analyzing OOMP2 density-related properties \cite{MP2-Molphys2009,OOMP2-HFCC-2010,sandhoefer2013derivation}.       
In the DFT community, the accuracy of electron density is a critical criterion for evaluating if a functional is good \cite{medvedev2017density}. A lot of functionals that give accurate energy-related properties have been found to have poor electron density \cite{medvedev2017density}. Recent studies have shown that double-hybrid functionals that are based on a mixing of a hybrid functional and a perturbative second-order correlation part (MP2) can yield an accurate prediction of both density- and energy-related properties \cite{su2018doubly, hait2018-dipole}. Note that the formal scaling of double-hybrid functionals is the same as MP2. Double-hybrid functionals with relaxed density are also ill-behaved for stretched bonds, giving unphysical density-related properties \cite{OODHFs-2013, hait2018-dipole, hait2018communication}. To resolve these issues, one can optimize orbitals of double-hybrid functionals like OOMP2 \cite{OODHFs-2013,OODHF-JPCA2016}. 

Recently we have developed a new self-consistent perturbation theory named one-body MP2 (OBMP2)\cite{OBMP2-JCP2013,tran2021improving}. The central idea of OBMP2 is the use of canonical transformation\cite{CT-JCP2006,CT-JCP2007,CT-ACP2007,CT-JCP2009,CT-JCP2010,CT-IRPC2010} followed by the cumulant approximation\cite{cumulant-JCP1997,cumulant-PRA1998,cumulant-CPL1998,cumulant-JCP1999} to derive an effective one-body Hamiltonian. The resulting OBMP2 Hamiltonian, as a {\it correlated} Fock operator, is a sum of the standard Fock operator and a one-body correlation MP2 potential. Molecular orbitals and orbital energies are relaxed in the presence of correlation by diagonalizing {\it correlated} Fock matrix. The double-excitation MP2 amplitudes are then updated using those new molecular orbitals and orbital energies, resulting in a self-consistency. Very recently, we have shown that the self-consistency of OBMP2 can resolve issues caused by the non-iterative nature of standard MP2 calculations for open-shell systems \cite{tran2021improving}. 

We explore the OBMP2 performance to predict electron density and related properties for open-shell systems in the present work. We consider three groups: atoms and their ions, main-row radicals, and halogen-bond dimers. We find that OBMP2 can outperform standard MP2 and is comparable to CCSD. It is important to emphasize that the computational scaling of OBMP2 is similar to that of MP2 ($N^5$) that is lower than CCSD one ($N^6$). Therefore, OBMP2 is highly promising for accurately predicting the properties of large molecules.    

\section{Theory }
Let us recap the OBMP2 theory whose formulation details are presented in Refs.~\citenum{tran2021improving} and ~\citenum{OBMP2-JCP2013}.
The OBMP2 approach was derived through the canonical transformation developed by Yanai and his coworkers\cite{CT-JCP2006,CT-JCP2007,CT-ACP2007,CT-JCP2009,CT-JCP2010,CT-IRPC2010}. In this approach, an effective Hamiltonian that includes dynamic correlation effects is achieved by a similarity transformation of the molecular Hamiltonian $\hat{H}$ using a unitary operator $e^{\hat{A}}$ :
\begin{align}
\hat{\bar{H}} = e^{\hat{A}^\dagger} \hat{H} e^{\hat{A}}
\label{Hamiltonian:ct}
\end{align}
with the anti-Hermitian excited operator $\hat{A} = - \hat{A}^\dagger$, and the molecular Hamiltonian in spin orbitals as
\begin{align}
  \hat{H} =  h^{p\sigma}_{q\sigma} \hat{a}_{p\sigma}^{q\sigma} + \tfrac{1}{2}g^{p\sigma r\sigma'}_{q\sigma s\sigma'}\hat{a}_{p\sigma r\sigma'}^{q\sigma s\sigma'}\label{eq:h1}
\end{align}
where $\left\{p, q, r, \ldots \right\}$ indices refer to general spatial orbitals and $\left\{\sigma, \sigma' \right\}$ refer to spin indices.
In OBMP2, the cluster operator $\hat{A}$ is modeled such that including only double excitation.
\begin{align}
  \hat{A} = \hat{A}_\text{D} = \tfrac{1}{2}T_{i\sigma j\sigma'}^{a\sigma b\sigma'}(\hat{a}_{i\sigma j\sigma'}^{a\sigma b\sigma'} - \hat{a}_{a\sigma b\sigma'}^{i\sigma j\sigma'}) \,, \label{eq:op1}
\end{align}
with the MP2 amplitude 
\begin{align}
  T_{i\sigma j\sigma'}^{a\sigma b\sigma'} =  \frac{g_{i\sigma j\sigma'}^{a\sigma b\sigma'} } { \epsilon_{i\sigma} + \epsilon_{j\sigma'} - \epsilon_{a\sigma} - \epsilon_{b\sigma'} } \,, \label{eq:amp}
\end{align}
where $\left\{i, j, k, \ldots \right\}$ indices refer to occupied spatial orbitals and
$\left\{a, b, c, \ldots \right\}$ indices refer to virtual spatial orbitals. $\epsilon_{i\sigma}$ is the orbital energy of the spin-orbital $i\sigma$. 

The OBMP2 Hamiltonian is defined as
\begin{align}
  \hat{H}_\text{OBMP2} = \hat{H}_\text{HF} + \left[\hat{H},\hat{A}_\text{D}\right]_1 + \tfrac{1}{2}\left[\left[\hat{F},\hat{A}_\text{D}\right],\hat{A}_\text{D}\right]_1.
 \label{eq:h2}
\end{align}
with
\begin{align}
  \hat{H}_\text{HF} = \hat{F} + C = f^{p\sigma}_{q\sigma} \hat{a}_{p\sigma}^{q\sigma} + \,\, C  \label{eq:h1hf}    
\end{align}
Where $\hat{F}$ is the Fock operator and $C$ is a constant.
In Eq.\ref{eq:h2}, commutators with the subscription 1, $[\ldots]_1$, involve one-body operators and constants that are reduced from many-body operators using the cumulant approximation\cite{cumulant-JCP1997,cumulant-PRA1998,cumulant-CPL1998,cumulant-JCP1999}. We eventually derive at the OBMP2 Hamiltonian as follows
\begin{align}
  \hat{H}_\text{OBMP2} = & \,\, \hat{H}_\text{HF} + \hat{V}_\text{OBMP2} \label{eq:h4}
\end{align}
where $\hat{V}_\text{OBMP2}$ is a correlated potential composing of one-body operators. The working expression are given in Refs.~\citenum{tran2021improving} and ~\citenum{OBMP2-JCP2013}. 
We rewrite $\hat{H}_\text{OBMP2}$ (Eqs. \ref{eq:h2} and \ref{eq:h4}) in a similar form to Eq. \ref{eq:h1hf} for $\hat{H}_\text{HF}$ as follows:
\begin{align}
  \hat{H}_\text{OBMP2} = & \hat{\bar{F}} + \bar{C} \label{eq:h5}
\end{align}
with $\hat{\bar{F}} =  \bar{f}^{p\sigma}_{q\sigma} \hat{a}_{p\sigma}^{q\sigma}$.
$\bar{f}^{p\sigma}_{q\sigma}$ is so-called {\it correlated} Fock matrix and written as
\begin{align}
\bar{f}^{p\sigma}_{q\sigma} &= f^{p\sigma}_{q\sigma} + v^{p\sigma}_{q\sigma}.
\end{align}

The perturbation matrix $v^{p\sigma}_{q\sigma}$ serves as the correlation potential altering the uncorrelated HF picture. 
We update the MO coefficients and energies by diagonalizing the matrix $\bar{f}^{p}_{q}$, leading to orbital relaxation in the presence of dynamic correlation effects. The OBMP2 method is implemented within a local version of PySCF\cite{pyscf-2018}. At the convergence, the OBMP2 electron density is evaluated using the $T_2$ amplitude (Eq.~\ref{eq:amp}) as in standard MP2.

\section{Results}
We first consider the atomic electron density that has been often used to examine the accuracy of DFT functionals\cite{medvedev2017density}. We measured the root-mean-square deviation (RMSD) relative to the CCSD electron density. For comparison, we also performed calculations of standard methods, including HF, MP2, and three DFT functional (B3LYP, HSE06, MN15-L) using pySCF\cite{pyscf-2018}. To see the effect of the orbital relaxation in OBMP2, we present MP2 results with the unrelaxed density that corresponds to the first OBMP2 iteration. 

\begin{table}[h!]
  \normalsize
  \caption{\label{tab:rms_dens} \normalsize Root mean square deviation (RMSD), normalized by $10^{-3}$ a.u, in atomic electron density. }
  \begin{tabular}{lcccccccccccccccc}
    \hline \hline		
      Atoms  &HF	&MP2$^a$	&OBMP2	&HSE06	&B3LYP	&MN15-L \\
    \hline
Al    &1.019	&0.842	&0.180	&2.623	 &2.986	&3.502 \\
B	  &0.890	&0.699	&0.427	&0.444	 &8.959	&10.119 \\
B$^+$ &1.019	&0.782	&0.600	&6.250	 &6.639	&6.558 \\
B$^-$ &0.827	&0.665	&0.341	&5.998	 &11.379   &13.979 \\
C     &0.877	&0.681	&0.413	&14.655 &8.927	&12.771 \\
C$^+$ &1.157	&0.889	&0.625	&17.833 &13.499	&10.542 \\
C$^-$ &0.717	&0.553	&0.122	&0.355	 &0.770	&3.032 \\
Cl	  &1.553	&1.108	&0.094	&8.204	 &10.806	&11.529 \\
F     &1.159	&0.782	&0.107	&5.396	 &22.571	&24.975 \\
N     &0.462	&0.336	&0.100	&0.442	 &0.922	&3.345 \\
N$^-$ &2.899	&2.810	&0.182	&4.861	 &9.812	&11.627 \\
N$^+$ &1.055	&0.806	&0.537	&21.990 &21.671	&23.387 \\
O     &0.833	&0.596	&0.117	&22.532 &22.569	&22.464 \\
O$^-$ &1.571	&1.056	&0.295	&19.904 &17.967	&9.971 \\
O$^+$ &0.382	&0.260	&0.080	&0.534	 &1.059	&3.464 \\
P     &1.359	&1.061	&0.079	&1.135	 &1.530	&3.088 \\
S	  &1.497	&1.126	&0.091	&6.159	 &9.685	&8.851 \\
Si	  &1.282	&1.033	&0.193	&5.896	 &6.957	&7.237 \\
\hline
MAD$^b$  &1.142	&0.894	&0.255	&8.067	 &9.928	 &10.580 \\
MAX$^b$  &2.899	&2.810	&0.625	&22.532 &22.571	&24.975 \\
\hline \hline
\end{tabular}\\
$^a$ The MP2 density matrix is not relaxed.  \\
$^b$ Mean absolute deviation (MAD) and maximum absolute deviation (MAX) relative to CCSD.
\end{table}

\begin{figure}[t!]
  \includegraphics[width=8cm,]{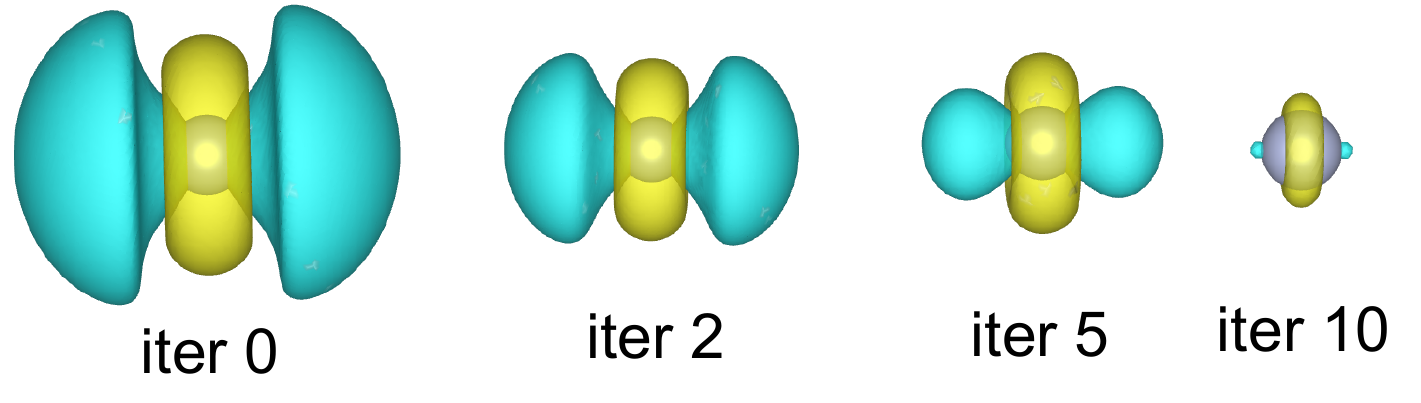}
  \caption{3D surfaces of the difference between OBMP2 and CCSD densities during the OBMP2 iteration for the N$^{-}$ system. The starting point of the self-consistency (iter = 0) correspond to standard MP2 without relaxation. The isosurface is 0.002.}
  \label{fig:dens-iter}
\end{figure}

Results are summarized in Table~\ref{tab:rms_dens}. All three DFT functionals have notoriously huge errors for atomic systems considered in this work. Although HF performs better than three DFT functionals, its errors are still quite large, especially for N$^-$. While MP2 can overall improve electron densities, it is still unable to give a satisfying error for N$^-$. When the electron density, evaluated directly using the $T_2$ amplitude (Eq.~\ref{eq:amp}), is relaxed via self-consistency in OBMP2, the errors are several times smaller than MP2 errors. In particular, as shown in Figure~\ref{fig:dens-iter}, the error for N$^{-}$ rapidly decreases with the OBMP2 iteration, indicating the importance of self-consistency.   

\begin{figure}[h!]
  \includegraphics[width=8cm,]{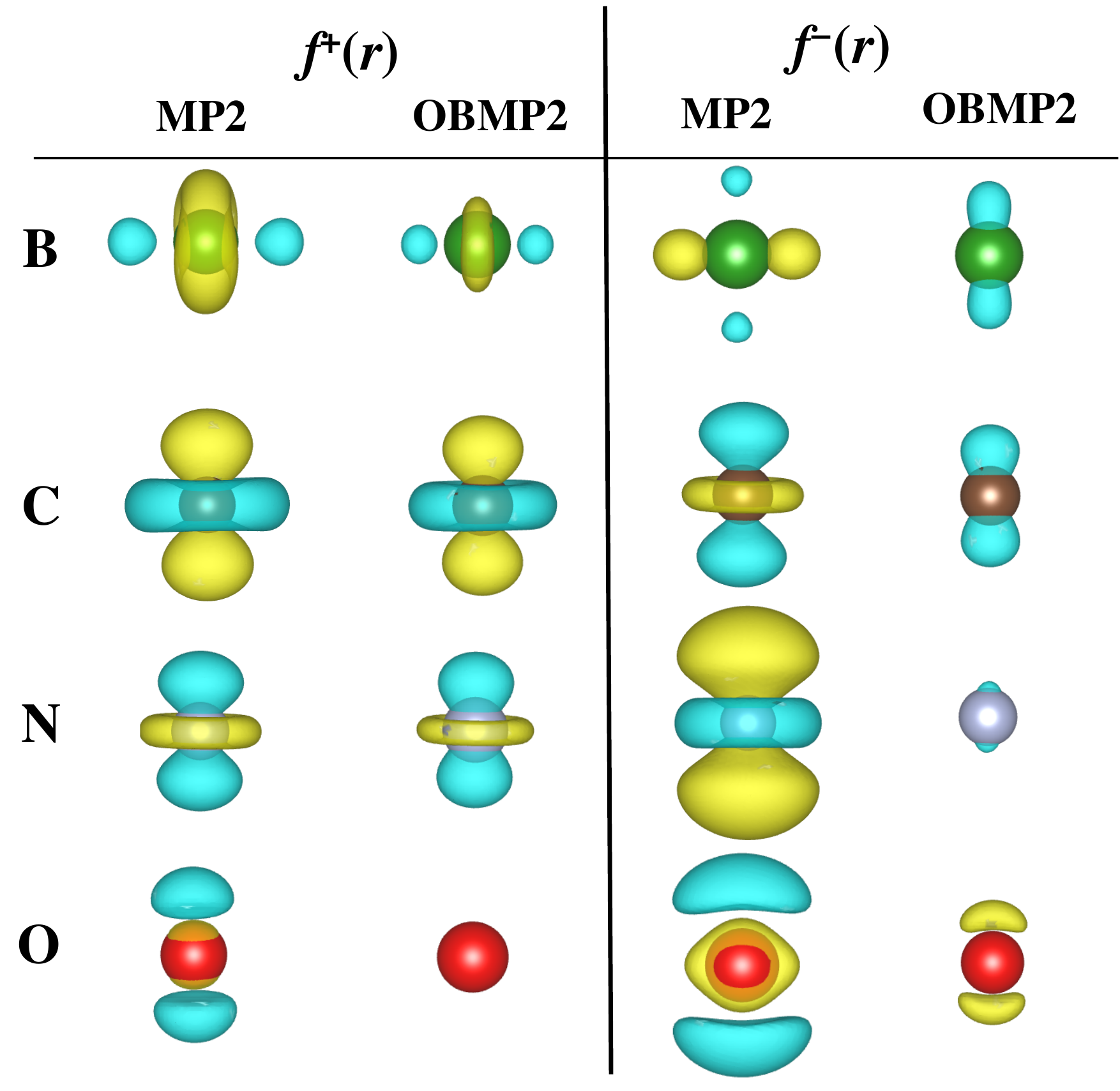}
  \caption{3D surfaces of errors in the Fukui functions $f^{+}(r)$ (left column) and $f^{-}(r)$ (right column) relative to the CCSD reference of four atoms B, C, N, O. The isosurface is 0.002.}
  \label{fig:fukui}
\end{figure}

We now analyze the Fukui functions for the removal of an electron $f^+(\bf r)$ and the addition of an electron $f^-(\bf r)$ that are defined as follows. 
\begin{align*}
    f^+(\bf r) = \rho^+({\bf r}) - \rho(\bf r) \\
    f^-(\bf r) = \rho({\bf r})   - \rho^-(\bf r) 
\end{align*}
Fukui functions$f^+(\bf r)$ and $f^-(\bf r)$ are the important regioselectivity indicators for chemical species prone to accepting and donating electrons in chemical reactions, respectively\cite{galvan1986fukui}. The reaction will take place where the Fukui functions can be found to have a large value. 3D surfaces of errors relative to the CCSD reference of MP2 and OBMP2 results are presented in Figure~\ref{fig:fukui}. Overall, except for the Fukui function $f^+(r)$ of C and N, OBMP2 errors are much smaller than standard MP2 errors and nearly vanish for $f^+(r)$ of O and $f^-(r)$ of N. The Fukui functions and their generalization have been widely used to predict molecular structure\cite{yanez2021kick}, we thus expect that OBMP2 will be an affordable and accurate approach to such an interesting topic.    

\begin{table*}[t!]
  \normalsize
  \caption{\label{tab:dipole} \normalsize Dipole moments of main-group radicals at equilibrium geometry in the aug-cc-pCVTZ basis set.}
  \begin{tabular}{lccccccccccccccccccccc}
    \hline \hline		
    \multirow{2}{*}{Radicals} &&\multirow{2}{*}{HF}  &\multicolumn{4}{c}{MP2} &&&&{DSD-}  &&\multirow{2}{*}{OOMP2} &&\multirow{2}{*}{OBMP2}	&&\multirow{2}{*}{CCSD}$^a$	&&\multirow{2}{*}{CCSD(T)}$^a$	\\
    \cline{4-6} 
    & &&unrelaxed  &&relaxed &&&&&BPEP86\\        
	\hline
AlH2    &&0.418   &0.407   &&0.434   &&&&&0.407   &&0.431   &&0.430   &&0.4069  &&0.399\\
BeH     &&0.271   &0.263   &&0.263   &&&&&0.247   &&0.260   &&0.262   &&0.2338  &&0.230\\
BH2     &&0.495   &0.497   &&0.498   &&&&&0.514   &&0.497   &&0.497   &&0.5006  &&0.500\\
BN      &&2.674   &2.674   &&2.211   &&&&&2.211   &&1.902   &&2.034   &&2.1593  &&2.000\\
BO      &&2.904   &2.738   &&2.468   &&&&&2.374   &&2.135   &&2.337   &&2.4028  &&2.293\\
BS      &&1.365   &1.239   &&1.123   &&&&&0.876   &&0.669   &&0.680   &&0.8451  &&0.730\\
CF      &&0.430   &0.508   &&0.735   &&&&&0.744   &&0.843   &&0.849   &&0.6275  &&0.692\\
CH      &&1.583   &1.567   &&1.506   &&&&&1.494   &&1.495   &&1.504   &&1.4423  &&1.416\\
CN      &&2.181   &2.133   &&2.305   &&&&&1.663   &&0.896   &&1.360   &&1.5683  &&1.410\\
FCO     &&1.283   &1.166   &&0.779   &&&&&0.782   &&0.653   &&0.685   &&0.8222  &&0.757\\
H2O-Al  &&4.741   &4.707   &&4.371   &&&&&4.386   &&4.322   &&4.382   &&4.4076  &&4.355\\
H2O-F   &&1.951   &1.912   &&1.935   &&&&&2.136   &&2.437   &&2.119   &&2.1011  &&2.173\\
LiN     &&7.357   &7.321   &&7.050   &&&&&7.050   &&7.017   &&7.087   &&7.1024  &&7.020\\
NF      &&0.325   &0.247   &&0.000   &&&&&0.041   &&0.183   &&0.078   &&0.007   &&0.069\\
NO      &&0.249   &0.145   &&0.031   &&&&&0.190   &&0.308   &&0.195   &&0.1136  &&0.137\\
NS      &&1.538   &1.520   &&2.076   &&&&&1.993   &&1.868   &&1.978   &&1.8406  &&1.839\\
OF      &&0.420   &0.356   &&0.225   &&&&&0.095   &&0.212   &&0.016   &&0.0653  &&0.016\\
PO      &&2.572   &2.364   &&1.864   &&&&&1.916   &&1.677   &&1.754   &&2.0908  &&1.935\\
PS      &&0.746   &0.695   &&0.464   &&&&&0.620   &&0.431   &&0.601   &&0.7487  &&0.633\\
SF      &&1.119   &1.024   &&0.833   &&&&&0.794   &&0.710   &&0.722   &&0.8862  &&0.808\\
SiH     &&0.242   &0.237   &&0.103   &&&&&0.130   &&0.099   &&0.114   &&0.1206  &&0.102\\\hline
MAD$^b$	&&0.330	&0.273	&&0.148	&&&&&0.065	&&0.124	&&0.058	&&0.068\\
MAX$^b$	&&0.772	&0.723	&&0.895	&&&&&0.254	&&0.514	&&0.181	&&0.160\\
\hline \hline
\end{tabular}\\
$^a$ Taken from Ref.\citenum{hait2018-dipole}. \\
$^b$ Mean absolute deviation (MAD) and maximum absolute deviation (MAX) relative to CCSD(T).
\end{table*}

We now consider the dipole moment that has been widely used as the simplest observable to examine the accuracy of new DFT functionals in predicting electron densities. For example, Hait and Head-Gordon \cite{hait2018-dipole} have recently carried out an extensive assessment of the performance of different DFT functionals belonging to different rungs of DFT Jacob's ladder. In the current work, we adopted 20 open-shell radicals from their work and employed the aug-cc-pCVTZ basis set. For comparison, we performed calculations for standard MP2 and a double-hybrid functional DSD-PBEP86 using ORCA \cite{orca-2020}, and orbital-optimized MP2 (OOMP2) using Psi4\cite{psi42020}. We adapted coupled-cluster results, CCSD and CCSD(T), from Ref.~\citenum{hait2018-dipole} and use CCSD(T) as the reference.

All results are summarized in Table~\ref{tab:dipole}. CCSD yields the result closest to the CCSD(T) reference. We can see that MP2 with unrelaxed density is slightly better than HF. While the density relaxation can help to reduce MP2 errors, they are still quite large. Surprisingly, OOMP2 errors are not much better than standard MP2 with relaxed density, whereas the double hybrid functional DSD-PBEP86 gives errors twice smaller than those of OOMP2. Overall, among MP2-based approaches, OBMP2 yields the smallest errors and is comparable to CCSD. 

\begin{figure*}[t!]
  \includegraphics[width=16cm,]{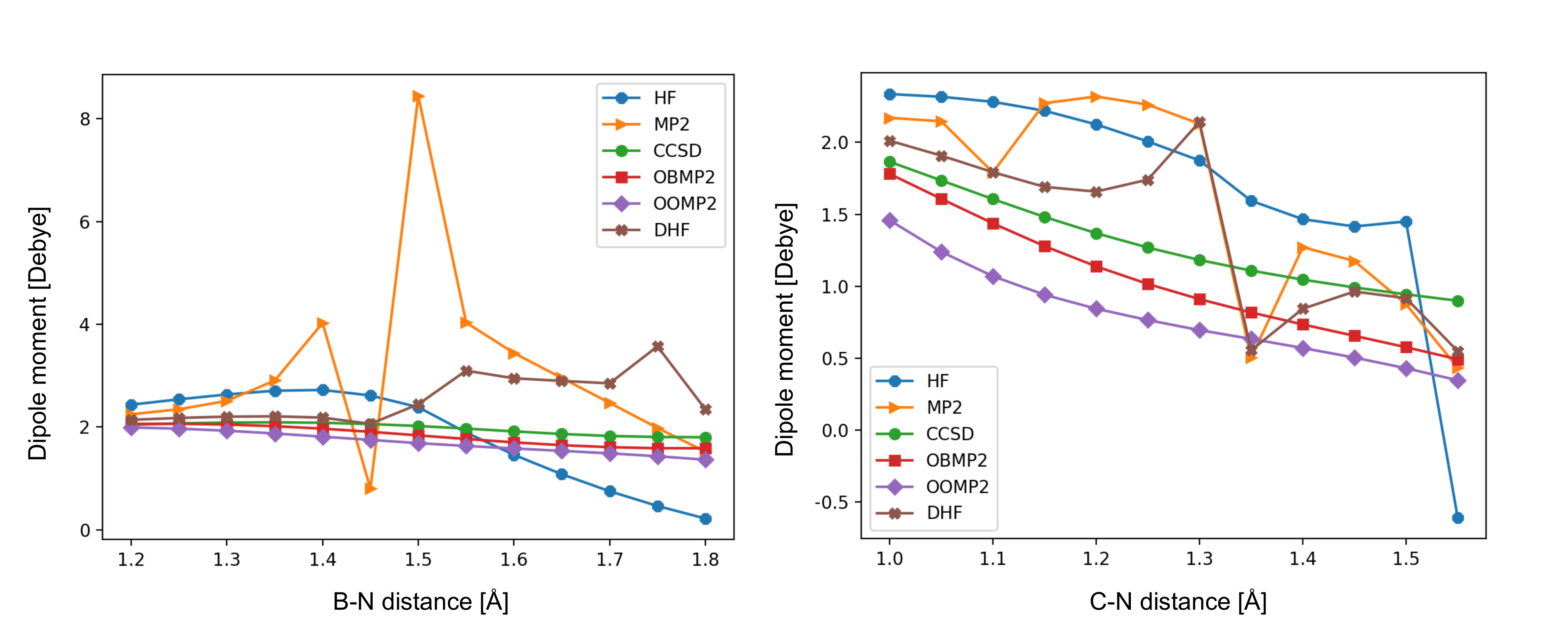}
  \caption{The change of dipole moments as stretching bond length for BN (left) and CN (right). DHF stands for double-hybrid functional DSD-PBEP86. Standard MP2 and DSD-PBEP86 densities are relaxed by solving the $Z-$vector equation for orbital response.}
  \label{fig:dip-R}
\end{figure*}

As previously discussed by some authors, standard MP2 and double-hybrid functionals with relaxed density give unphysical jumps of dipole moments in the bond-stretched regime. This happens due to the singularity of the inverse of the orbital Hessian matrix used for evaluating the orbital response of MP2 density matrix\cite{MP2-Molphys2009}. Head-Gordon and some other authors have shown that orbital optimization is crucial to avoid unphysical jumps of dipole moment \cite{hait2018communication}. Thus, it is interesting to examine how OBMP2 dipole moments behave as stretching molecules. In Figure~\ref{fig:dip-R}, we plot the change of dipole moments of BN and CN as functions of bond length. For both radicals, it is clear that MP2 and DSD-PBEP86 dipole moments with relaxed densities exhibit unphysical jumps. OOMP2 and OBMP2, on the contrary, have continuous curves. 

\begin{table*}[t!]
  \normalsize
  \caption{\label{tab:hal-charge} \normalsize Atomic Mulliken charge of halogen-bond systems. }
  \begin{tabular}{llllccccccccccccc}
    \hline \hline		
    Systems     &&&\,\,HF	&&MP2	&&\,\,OBMP2	&&\,\,HSE06	&&\,\,B3LYP	&&\,\,MN15-L &&\,\,CCSD\\
    \hline
    FBr-Br     &F      &&$-0.5540$ &&$-0.5408$ &&$-0.4573$ &&$-0.4767$ &&$-0.4639$ &&$-0.4780$ &&$-0.4654$ \\
           &Br1    &&$\,\,\,\,0.5068$ &&$\,\,\,\,0.4921$ &&$\,\,\,\,0.4036$ &&$\,\,\,\,0.5009$ &&$\,\,\,\,0.4589$ &&$\,\,\,\,0.4643$ &&$\,\,\,\,0.4165$ \\
           &Br2    &&$\,\,\,\,0.0472$ &&$\,\,\,\,0.0486$ &&$\,\,\,\,0.0537$ &&$-0.0242$ &&$\,\,\,\,0.0050$ &&$\,\,\,\,0.0138$ &&$\,\,\,\,0.0489$ \\
FBr-Cl     &F      &&$-0.5375$ &&$-0.5245$ &&$-0.4431$ &&$-0.4608$ &&$-0.4485$ &&$-0.4597$ &&$-0.4503$ \\
           &Br     &&$\,\,\,\,0.5250$ &&$\,\,\,\,0.5114$ &&$\,\,\,\,0.4284$ &&$\,\,\,\,0.5403$ &&$\,\,\,\,0.4917$ &&$\,\,\,\,0.5194$ &&$\,\,\,\,0.4377$ \\
           &Cl     &&$\,\,\,\,0.0125$ &&$\,\,\,\,0.0131$ &&$\,\,\,\,0.0147$ &&$-0.0794$ &&$-0.0432$ &&$-0.0597$ &&$\,\,\,\,0.0126$ \\
FBr-OH     &F      &&$-0.5556$ &&$-0.5428$ &&$-0.4607$ &&$-0.4834$ &&$-0.4698$ &&$-0.4838$ &&$-0.4682$ \\
           &Br     &&$\,\,\,\,0.5441$ &&$\,\,\,\,0.5306$ &&$\,\,\,\,0.4416$ &&$\,\,\,\,0.5083$ &&$\,\,\,\,0.4621$ &&$\,\,\,\,0.4261$ &&$\,\,\,\,0.4523$ \\
           &O      &&$-0.2903$ &&$-0.2830$ &&$-0.2696$ &&$-0.2754$ &&$-0.2555$ &&$-0.3659$ &&$-0.2694$ \\
           &H      &&$\,\,\,\,0.3018$ &&$\,\,\,\,0.2952$ &&$\,\,\,\,0.2887$ &&$\,\,\,\,0.2506$ &&$\,\,\,\,0.2632$ &&$\,\,\,\,0.4236$ &&$\,\,\,\,0.2854$ \\
FCl-Br     &F      &&$-0.4074$ &&$-0.3956$ &&$-0.3305$ &&$-0.3618$ &&$-0.3438$ &&$-0.3566$ &&$-0.3363$ \\
           &Cl     &&$\,\,\,\,0.3834$ &&$\,\,\,\,0.3707$ &&$\,\,\,\,0.2981$ &&$\,\,\,\,0.3360$ &&$\,\,\,\,0.3079$ &&$\,\,\,\,0.3262$ &&$\,\,\,\,0.3076$ \\
           &Br     &&$\,\,\,\,0.0240$ &&$\,\,\,\,0.0249$ &&$\,\,\,\,0.0323$ &&$\,\,\,\,0.0258$ &&$\,\,\,\,0.0359$ &&$\,\,\,\,0.0304$ &&$\,\,\,\,0.0287$ \\
FCl-Cl     &F      &&$-0.3931$ &&$-0.3814$ &&$-0.3179$ &&$-0.3472$ &&$-0.3291$ &&$-0.3405$ &&$-0.3236$ \\
           &Cl1    &&$\,\,\,\,0.3857$ &&$\,\,\,\,0.3738$ &&$\,\,\,\,0.3066$ &&$\,\,\,\,0.3463$ &&$\,\,\,\,0.3156$ &&$\,\,\,\,0.3234$ &&$\,\,\,\,0.3141$ \\
           &Cl2    &&$\,\,\,\,0.0074$ &&$\,\,\,\,0.0077$ &&$\,\,\,\,0.0113$ &&$\,\,\,\,0.0009$ &&$\,\,\,\,0.0135$ &&$\,\,\,\,0.0172$ &&$\,\,\,\,0.0095$ \\
FCl-OH     &F      &&$-0.4092$ &&$-0.3975$ &&$-0.3333$ &&$-0.3627$ &&$-0.3440$ &&$-0.3547$ &&$-0.3388$ \\
           &Cl     &&$\,\,\,\,0.4080$ &&$\,\,\,\,0.3960$ &&$\,\,\,\,0.3232$ &&$\,\,\,\,0.3472$ &&$\,\,\,\,0.3223$ &&$\,\,\,\,0.3102$ &&$\,\,\,\,0.3317$ \\
           &O      &&$-0.2877$ &&$-0.2807$ &&$-0.2667$ &&$-0.2300$ &&$-0.2309$ &&$-0.3704$ &&$-0.2668$ \\
           &H      &&$\,\,\,\,0.2890$ &&$\,\,\,\,0.2822$ &&$\,\,\,\,0.2768$ &&$\,\,\,\,0.2456$ &&$\,\,\,\,0.2526$ &&$\,\,\,\,0.4149$ &&$\,\,\,\,0.2738$ \\
\hline
MAD$^a$  &&&\,\,\,\,0.0525	&&\,\,\,\,0.0433	&&\,\,\,\,0.0059	&&\,\,\,\,0.0356	&&\,\,\,\,0.0172	&&\,\,\,\,0.0446 \\
MAX$^a$  &&&\,\,\,\,0.0918	&&\,\,\,\,0.0783	&&\,\,\,\,0.0129	&&\,\,\,\,0.1025	&&\,\,\,\,0.0557	&&\,\,\,\,0.1411 \\
\hline \hline
\end{tabular} \\
$^a$Mean absolute deviation (MAD) and maximum absolute deviation (MAX) relative to CCSD.
\end{table*}

The last test set we consider is of halogen-bonding dimmers. In these systems, the charge transfer between monomers importantly contributes to the bonding mechanism \cite{cavallo2016halogen}. Here, we consider six open-shell dimmers of FX--Y (X = Br, Cl, and Y = Br, Cl, OH) whose geometries were adopted from Ref.~\citenum{bandyopadhyay2020components}. We use the aug-cc-pVTZ basis set and CCSD as a reference. WE did not relax the density matrix of standard MP2 to show the importance of orbital optimization in OBMP2. We also carried out three DFT functional calculations for comparison, including B3LYP, HSE06, and MN15-L. All these calculations were done using pySCF\cite{pyscf-2018}.

\begin{figure}[h!]
  \includegraphics[width=8cm,]{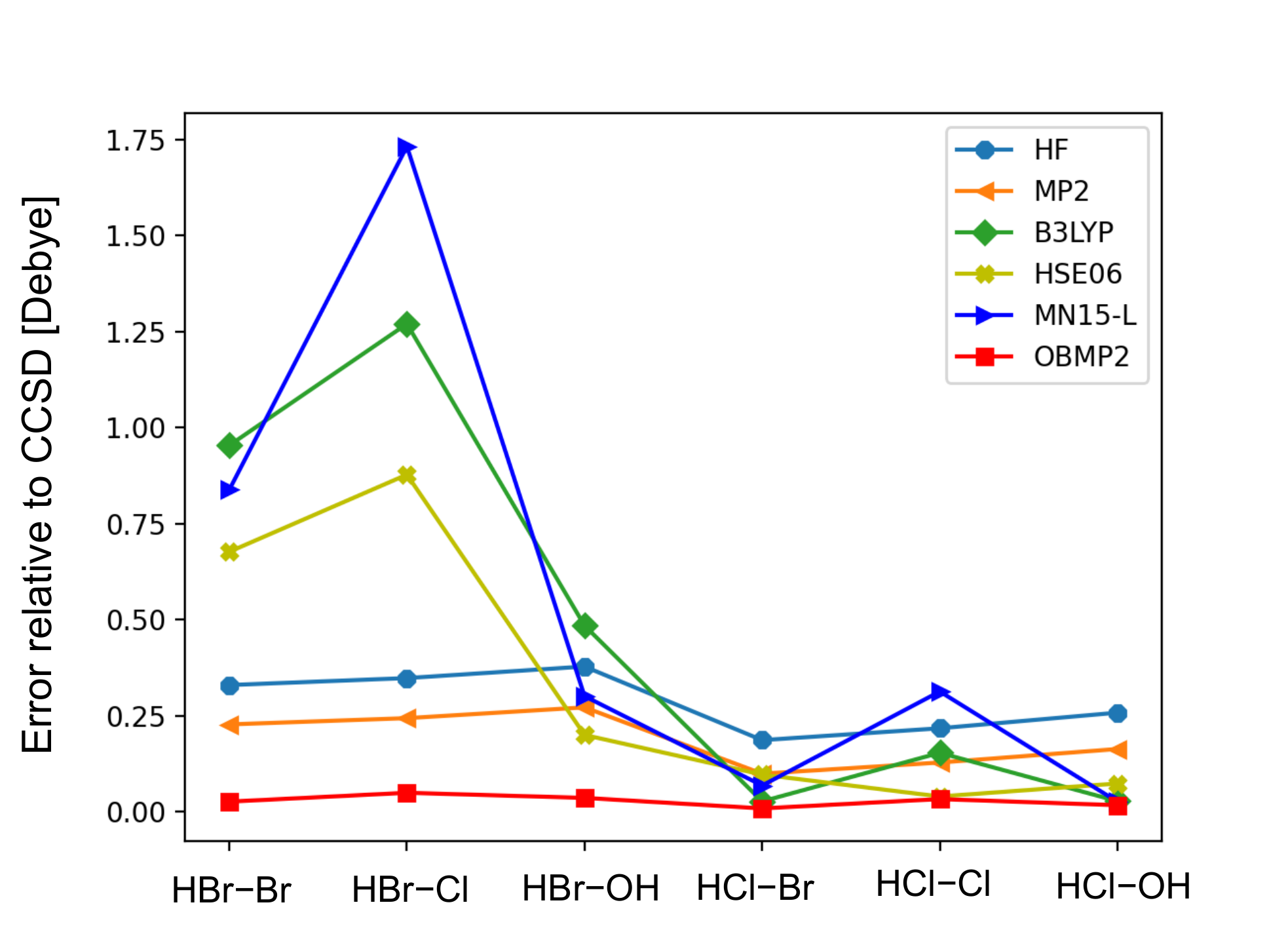}
  \caption{Errors relative to the CCSD reference in dipole moments of halogen-bond systems evaluated using some wavefunction and DFT methods.}
  \label{fig:hal-dip}
\end{figure}

Table~\ref{tab:hal-charge} shows atomic Mulliken charges of the six dimmers. We can see that MP2 with unrelaxed density is slightly better than HF. Among three DFT functionals, B3LYP performs best, and its errors are even smaller than standard MP2. Thanks to the orbital optimization {\it via} self-consistency, OBMP2 dramatically improves upon MP2 results. Compared with other methods considered here, its errors relative to the CCSD reference are the smallest. Figure~\ref{fig:hal-dip} depicts errors in dipole moments relative to the CCSD reference. It is not surprising that the DFT performance is system-dependent. While their errors are small for HCl--Br and HCl--OH, they are large for other systems, particularly HBr--Cl. HF and standard MP2 performance are quite stable, but their errors are still large. OBMP2 outperforms standard MP2 and DFT and yields the smallest errors with a maximum of 2.3\% for HBr--Cl.  

\section{Conclusion}
In summary, we have presented the performance of our recently-developed OBMP2 method to predict electron density and related properties of open-shell systems. We have considered three groups: atoms and their ions, main-group radicals, and halogen-bonding dimmers. We compared the OBMP2 performance against standard methods, including HF, MP2, DFT, and coupled clusters. We have found that OBMP2 yields result closest to CCSD in all cases considered in this work. It is worth reminding that the formal scaling of OBMP2 is the same as standard MP2 ($N^5$), which is lower than CCSD ($N^6$). OBMP2 is thus believed to be an effective method for generating accurate electron densities. Further speeding up OBMP2 by employing local and density-fitting approximations is in progress.   

\section*{Acknowledgments}
This work is supported by the Vietnam Academy of Science and Technology (VAST) through the VAST Program for Young Researchers under the grant number DLTE00.02/22-23. 
We performed all calculations on the High-Performance Computing system at the Center for Informatics and Computing (CIC), VAST.  
\bibliography{main}
\end{document}